\documentclass[10pt]{aimeta2013}

\usepackage{amsfonts}
\usepackage{subfigure}
\usepackage{hyperref}
\usepackage{graphicx}
\usepackage{color}
 \usepackage{epsfig}
\usepackage{amssymb}
\usepackage{amsmath}
\usepackage{amsthm}
\usepackage{booktabs}
\usepackage{verbatim}

\title{An extended SMLD approach for presumed probability density function in flamelet combustion model}

\author{Alessandro Coclite$^{1, \dag,*}$, Giuseppe Pascazio$^{1,*}$, Pietro De Palma$^{1,*}$, Luigi Cutrone$^{2,*}$}

\address{{\em\bigskip $^{1}$Dipartimento di Meccanica, Matematica e Management (DMMM), \\

Politecnico di Bari,  Via Re David 200, 70125, Bari Italy
\\ E-mail: a.coclite@poliba.it, pascazio@poliba.it, depalma@poliba.it. 

\bigskip
$^{2}$Centro Italiano Ricerche Aerospaziali (CIRA), Via Maiorise 81043, Capua, Italy,
\\ E-mail: l.cutrone@cira.it

\bigskip
$^{*}$Centro di Eccellenza in Meccanica Computazionale (CEMEC), Via Re David 200, 70125, Bari Italy\\

\bigskip
$^{\dag}$Corresponding author\\
}}

\abstract{Flamelet Progress Variable approach, Non-premixed combustion, Statistically Most Likely Distribution.
\\[17pt]
SUMMARY. This paper provides an extension of the standard flamelet progress variable (FPV) approach for turbulent combustion, applying the statistically most likely distribution (SMLD) framework to the joint PDF of the mixture fraction, Z, and the progress variable, C. In this way one does not need to make any assumption about the statistical correlation between Z and C and about the behaviour of the mixture fraction, as required in previous FPV models. In fact, for state-of-the-art models, with the assumption of very-fast-chemistry,Z is widely accepted to behave as a passive scalar characterized by a $\beta$-distribution function. Instead, the model proposed here, evaluates the most probable joint distribution of  Z and C without any assumption on their behaviour and provides an effective tool to verify the adequateness of widely used hypotheses, such as their statistical independence. 
The model is validated versus three well-known test cases, namely, the Sandia flames. The results are compared with those obtained by the standard FPV approach, analysing the role of the PDF functional form on turbulent combustion simulations.
}

\begin{document}

\section{INTRODUCTION}

Turbulent combustion is a formidable multi-scale problem, where the interaction between chemical kinetics, molecular, and turbulent transport occurs over a very wide range of length and time scales. The numerical simulation of such phenomena with detailed chemistry is today prohibitive, so that a reduction model is often employed to simplify the reaction mechanisms and cut down the computational costs. Therefore, different approaches have been proposed to address this problem, such as the reduction of the chemical scheme in intrinsic low dimensional manifolds~(ILDM)~\cite{maas}; the flamelet-based approaches such as the flamelet-progress variable~(FPV)~\cite{pierce,piercemoin2004} or flame prolongation of ILDM~(FPI)~\cite{laminarhydrogen}; and Flamelet Generated Manifolds approach~(FGM)~\cite{oijen}. 
Our interest is devoted here to diffusive, either partially premixed or non-premixed, flames which constitute a specific class of combustion problems where fuel and oxidizer enter separately into the combustion chamber. 
Non-premixed flames can be characterized by a local balance between diffusion and reaction~\cite{peters} and their structure can be described by a conserved scalar, the mixture fraction. A diffusive flame can be viewed as an ensemble of thin locally one-dimensional structures embedded within the flow field. Each element of the flame front can then be described as a small laminar flame, also called {\it flamelet}. 
In this paper we focus on FPV approach for turbulent non-premixed flames. The FPV approach is based on the use of only two degrees of freedom, namely, the mixture fraction, Z, and the progress variable, C, that are employed to map all of the thermodynamic quantities involved in the process.
For the case of a turbulent flame one needs to define a probability density function (PDF) to compute the Favre average of the thermo-chemical quantities. The accuracy of the model depends on the
definition of such a distribution, whose properties are critical due to the poor knowledge of the two independent variables behaviour. The aim of this work is to provide an extension of the standard FPV model for turbulent combustion, applying the statistically most likely distribution (SMLD)~\cite{pope} approach to the joint PDF of $Z$ and $C$. The rational behind the definition of such a PDF is based on the reconsideration of the statistical independence hypothesis of $Z$ and $C$. It can be shown that assuming the steady laminar flamelet equation to parametrize all of the thermo-chemical quantities as functions of $Z$ and $C$, is equivalent to suppose the statistical independence of the two scalars~\cite{peters84}; but it is also true that the steady laminar flamelet equation is still valid, even if $Z$ and $C$ are dependent, as long as their statistical behaviour is accurately presumed in the joint PDF~\cite{ihmeal2005}.  
Four PDF models are considered and their role in the evaluation of non-premixed flames is analysed. This is assessed in the third section, where the numerical results obtained in the simulation of the Sandia flames~\cite{sandia} are discussed. The paper closes with summary and conclusions.

\section{THE MODEL}

\subsection{The flamelet approach}
The FPV model proposed by Pierce and Moin~\cite{pierce,piercemoin2004} is used in this work to evaluate all of the thermo-chemical quantities involved in the combustion process. This approach is based on the parametrization of the generic quantity, $\phi$, in terms of two variables, the mixture fraction $Z$ and the progress variable $C$:
\begin{equation}
\label{phi}
{\phi=F_\phi(Z,C)}.
\end{equation}
Equation \eqref{phi} is taken as the solution of the steady laminar flamelet equation:
\begin{equation}
\label{slfe}
{-\rho\frac{\chi}{2}\frac{\partial^2 \phi}{\partial Z^2}=\dot\omega_\phi},
\end{equation}
where $\chi$ is the scalar dissipation rate modeled in terms of the molecular diffusivity of $Z$, $D_Z$, $\chi=2D_Z(\nabla Z)^2$; $\rho$ is the density; $\dot\omega_\phi$ is the source term related to $\phi$. Each solution of equation~\eqref{slfe} is a flamelet and the solution variety over $\chi=\chi_{st}$ is called S-curve. 
From equation \eqref{phi} one can obtain the Favre-averages of $\phi$ using the definitions:
\begin{equation}
\label{media}
{\widetilde\phi=\int\int F_\phi(Z,C)\widetilde{P}(Z,C)dZ dC},
\end{equation}
\begin{equation}
\label{varianza}
{\widetilde{\phi''^2}=\int\int (F_\phi(Z,C)-\widetilde\phi)^2\widetilde{P}(Z,C)dZdC},
\end{equation}
where $\widetilde P(Z,C)$ is the density-weighted PDF,
\begin{equation}
{\widetilde P(Z,C)=\frac{\rho P(Z,C)}{\overline{\rho}}},
\end{equation}
$P(Z,C)$ is the joint PDF and $\overline{\rho}$ is the Reynolds-averaged density. As usual, $\phi$ can be decomposed as:
\begin{equation}
\phi=\widetilde \phi+ \phi'' \, ,
\qquad
\widetilde \phi = \frac{\overline{\rho \phi }}{\overline{\rho}} \, ,
\qquad
\rho=\overline{\rho} +\rho' \, ,
\end{equation}
where $\phi''$ and $\rho'$ are the fluctuations.
This ensures that the filtering process does not alter the form of the conservation laws.\\
The choice of the PDF plays a crucial role in the definition of the model, being a compromise between computational costs and accuracy level. In this respect, this paper provides an extension of the standard FPV turbulent combustion model combined with a RANS equation solver~\cite{luigi}. The final aim of this research is to study the influence of the different PDFs in the simulation of non-premixed turbulent combustion.
\subsection{\em Presumed probability density function modeling}

In order to investigate the role of the presumed PDF one can, first of all, use the Bayes' theorem and take the PDF as the product between the marginal PDF of $Z$ and the conditional PDF of $C|Z$:
\begin{equation}
\label{bayes}
{\widetilde{P}(Z,C)=\widetilde P(Z)\widetilde P(C|Z)}.
\end{equation}
Therefore, one has to presume the functional shape of such PDFs. Let us consider the marginal PDF, $\widetilde P(Z)$. It has been shown, by several authors, that in the limit of infinitely fast chemistry, implying the zero thickness limit of the reaction zone, the solution of the one-dimensional non-premixed flame (Burke-Schumann solution) is correctly reproduced by computing only one passive scalar, namely, the mixture fraction, see, e.g.,~\cite{peters}, whose statistical behavior can be estimated by a $\beta$~distribution~\cite{cook,jimenez,wall}. 
In the first three models discussed in this work, the $\beta$-distribution is employed for $\widetilde P(Z)$. Moreover, to presume the functional shape of the distribution of a reacting scalar, one needs to make some constitutive hypotheses. To simplify the problem, in this work we assume the statistical independence of $Z$ and $C$ for the first three models, so that, eq.~(\ref{bayes}) reads $\widetilde P(Z,C)=\widetilde P(Z)\widetilde P(C)$, namely $C=C|Z$. The most widely used hypothesis (model~A), implying a great simplification in the theoretical framework, consists in assuming that $\widetilde P(C)$ is modeled by a Dirac distribution, providing only one solution of equation~\eqref{slfe} for each chemical state. With this criterion the Favre-average of a generic thermo-chemical quantity is given by:
\begin{equation}
\label{phidelta}
{\widetilde\phi=\int\int F_\phi(Z,C)\widetilde \beta(Z)\delta(C-\widetilde{C})dZdC=\int F_\phi(Z,\widetilde C)\widetilde \beta(Z)dZ}.
\end{equation} 
Therefore, the resulting model employs only three additional transport equations (for $\widetilde Z$, $\widetilde{Z''^2}$ and $\widetilde C$) to evaluate all thermo-chemical quantities in the flow thus avoiding the expensive solution of a transport equation for each chemical species. The statistical behaviour of $Z$ and $C$ is strongly affected by the hypotheses posed to build model~A. In fact, it is well known that a reactive scalar~\cite{ihmea}, such as $C$, depends on a combination of solutions of equation~\eqref{slfe} for each chemical state and therefore its PDF cannot be accurately approximated by a Dirac distribution.\\
Thereby, the second model (model~B) is designed by assuming that $Z$ and $C$ are distributed in the same way, namely, using a $\beta$-distribution, thus giving the joint PDF: 
\begin{equation}
\label{betabeta}
{\widetilde P(Z,C)=\widetilde\beta(Z)\widetilde\beta(C)}.
\end{equation}
This does not allow the simplification seen before and, consequently, the model requires the evaluation of an additional transport equation for $\widetilde{C''^2}$. \\
Moreover, the probability distribution of a reacting scalar is often multi-modal, unlike the $\beta$ function, and its functional form depends on the turbulence-chemistry interaction. Therefore, one can think about a distribution built considering, as constraints, the only available informations, namely the value of $\widetilde Z$, $\widetilde{Z''^2}$, $\widetilde C$ and $\widetilde{C''^2}$. The third model (model~C) is obtained evaluating the conditional PDF as the statistically most likely distribution (SMLD)~\cite{ihmea}. It can be shown that if one knows only its first three moments, the PDF can be evaluated using ``Laplace's principle of insufficient reason''~\cite{pope}. The technique is developed following the statistical mechanics arguments presented by Heinz~\cite{heinz}. Relying on the knowledge of the first three moments of $\widetilde P(C)$, a unique measure, $S$, of the predictability of a thermodynamic state can be defined. $S$ is an entropy function depending on $\widetilde P(C)$, $S=S(\widetilde P(C))$~\cite{shannon} that can be thought of as the Boltzmann's entropy:
\begin{equation}
{S=-\int \widetilde P(C) \ln\Bigl(\frac{\widetilde P(C)}{Q(C)}\Bigr) dC},
\end{equation}
\noindent where $Q(C)$ is a bias density function to integrate information when no moments are known. In this paper the form of $Q(C)$ proposed by Pope~\cite{ihmeb} is assumed. The goal is to construct a PDF that maximizes the entropy $S$. Following the Lagrangian optimization approach, the functional $S^*$ is defined by involving the constraints on the moments:
\begin{equation}
{S^*=-\int dC\Bigl\{ \widetilde P(C) \ln\Bigl(\frac{\widetilde P(C)}{Q(C)}\Bigr)+\sum_{n=1}^2 \mu_n C^n \widetilde P(C)-\frac{\widetilde P(C)}{Q(C)} \Bigr\}}.
\end{equation}
In the above equation $\mu_n$ are the Lagrange's multipliers while the last fraction term is introduced to normalize $\widetilde P(C)$. 
The expression for $\widetilde P(C)$, obtained evaluating the maximum of $S^*$, reads: 
\begin{equation}
\label{smld}
{\widetilde P(C)=\frac{1}{\mu_0}\exp\Bigl\{-\sum_{n=1}^2 \frac{\mu_n}{n}(C-\widetilde C)^n \Bigr\}},
\end{equation}
where:
\begin{eqnarray}
\mu_0&=&\int_0^1 dC \widetilde P(C),\\
-\mu_1&=&\int_0^1 dC \partial_C (\widetilde P(C))=\widetilde P(1)-\widetilde P(0),\\
1-\mu_2\widetilde{C''^2}&=&\int_0^1 dC \partial_C[(C-\widetilde C)\widetilde P(C)]=\widetilde P(1)-\widetilde C\mu_1,
\end{eqnarray}
since $Z$ and $C$ are bounded in $[0,1]$.

At this point the model still needs an additional assumption to be closed. Here we assume that the first and the last point of $\widetilde P(C)$ are equal to the first and last points of $\beta(C)$ evaluated with the given values of the mean and variance:
\begin{equation}
{\widetilde P(1;\widetilde C,\widetilde{C''^2})=\widetilde\beta(1;\widetilde C,\widetilde{C''^2})},\\
{\ \ \ \widetilde P(0;\widetilde C,\widetilde{C''^2})=\widetilde\beta(0;\widetilde C,\widetilde{C''^2})}.
\end{equation}
This assumption does not affect the multi-modal nature of the distribution, but simplifies the model implementation (there is no need to evaluate the roots of a non-linear system).
The major advantage of the SMLD approach over conventionally employed presumed PDF closure models is that it provides a systematic framework to incorporate an arbitrary number of moment information.
It is noteworthy that, since $C$ is used instead of $C|Z$ as argument of $\widetilde P$, also this model assumes statistical independence of $Z$ and $C$.\\
In order to overcome the limits of the models described above, one should avoid the use of any hypothesis, establishing a more general design framework.
Thereby, even considering the solutions of equation \eqref{slfe}, one can properly assess the statistical correlation of $Z$ and $C$ with an accurate estimation of the joint PDF~\cite{ihmeal2005}. Our proposal (model~D) is to apply the SMLD framework directly to the joint distribution. In this way, one does not need any assumption on the statistical correlation between $Z$ and $C$ and can evaluate the most probably distribution of $Z$ without the very fast chemistry hypothesis. 
Let us suppose the knowledge of the first three moments of the joint probability $\widetilde P(\vec x)$, were $\vec x=(Z,C)^T$.
Using the statistical arguments of model~C, the following two dimensional PDF is obtained:
\begin{multline}
\label{smld2}
\widetilde P_{SML,2}(Z,C)= \frac{1}{\mu_0}\exp\Bigl\{-\Bigl[\mu_{1,1} (Z-\widetilde Z)+\mu_{1,2}(C-\widetilde C)\Bigr]\\-\frac{1}{2}\Bigl[\mu_{2,11}(Z-\widetilde Z)^2+\mu_{2,12}(Z-\widetilde Z)(C-\widetilde C)\\+\mu_{2,21}(C-\widetilde C)(Z-\widetilde Z)+\mu_{2,22}(C-\widetilde C)^2 \Bigl]
\Bigr\}.
\end{multline}
Since $\widetilde P(\vec{x})$ is a function of two variables, $\mu_0$ is a scalar, $\vec{\mu_1}$ is a two component vector, and $\overleftrightarrow{\mu_2}$ is a square matrix of rank two:
\begin{eqnarray}
\label{moltiplicatori}
\mu_0&=&\int_0^1 d\vec x \widetilde P_{SML,2}(\vec x) ,\\
\label{moltiplicatori1}
-\mu_{1,i}&=&\int_0^1 d\vec x\partial_{x_i} \widetilde P_{SML,2}(\vec x)=\beta(1;\widetilde\xi_i,\widetilde{\xi_i''^2})-\beta(0;\widetilde \xi_i,\widetilde{\xi_i''^2}) ,\\
\label{moltiplicatori2}
\delta_{kl}-\mu_{2,kn}\ \widetilde{\xi'_n\xi'_l}&=&\int_0^1 d\vec x \partial_{x_k}((x_l-\widetilde\xi_l)\widetilde P_{SML,2}(\vec x))=\beta(1;\widetilde \xi_k,\widetilde{\xi'_k\xi'_l})-\widetilde\xi_k\mu_{1,l} .
\end{eqnarray} 
It is interesting to note that, applying the Bayes' theorem to $\widetilde P_{SML,2}(Z,C)$ and assuming $\beta$-distribution for the marginal PDF, this model automatically turns into model~C. 
In fact, one can observe that:
\begin{equation}
{\widetilde P_{SML,2}(Z,C)= \widetilde P(Z)\widetilde P_{SML,2}(C|Z)=\widetilde\beta(Z)\widetilde P_{SML,2}(C|Z)};
\end{equation}
in this case the first multiplier $\mu_0$ is still given by equation \eqref{moltiplicatori}; the second and the third ones, $\vec{\mu_1}$ and $\overleftrightarrow{\mu_2}$, are reduced to a scalar.

\subsection{\em Turbulent FPV transport equations}
\vspace{3mm}
For the case of a turbulent flame, equation~\eqref{phi} must be written in terms of the Favre averages of $Z$ and $C$ and in terms of their variance. Using model~A one can tabulate all chemical quantities in terms of $\widetilde Z$, $\widetilde{Z^{''2}}$ and $\widetilde C$ because of the properties of the $\delta$-distribution. On the other hand, models~B, C and D express $\phi$ in terms of $\widetilde{C^{''2}}$ too and therefore they need to evolve a transport equation also for $\widetilde{C^{''2}}$. 
The transport equations read:
\begin{eqnarray}
\label{zmean}
 \partial_t(\overline{\rho}\widetilde{Z})+\vec\nabla\cdot(\overline{\rho}\widetilde{\vec u}\widetilde{Z})&=&
\vec\nabla\cdot\Bigl[\bigl( D+
D_{\widetilde{Z}}^t\bigr)\overline{\rho}
\vec\nabla\widetilde{Z}\Bigr],\\
\label{zvar}
\partial_t(\overline{\rho}\widetilde{Z''^2})+\vec\nabla\cdot(\overline{\rho}\widetilde{\vec u}\widetilde{Z''^2})&=&
\vec\nabla\cdot\Bigl[\bigl( D+D_{\widetilde{Z''^2}}^t\bigr)\overline{\rho}\vec\nabla\widetilde{Z''^2}\Bigr]-
\nonumber\\&-&\overline{\rho}\widetilde{\chi}+2\overline{\rho}D_{\widetilde Z}^t(\vec\nabla\widetilde{Z})^2,\\
\label{cmean}
\partial_t(\overline{\rho}\widetilde{C})+\vec\nabla\cdot(\overline{\rho}\widetilde{\vec u}\widetilde{C})&=&
\vec\nabla\cdot\Bigl[\bigl( D+D_{\widetilde{C}}^t\bigr)\overline{\rho}\vec\nabla\widetilde{C}\Bigr]+\overline{\rho}\overline{\dot\omega_C},\\
\label{cvar}
\partial_t(\overline{\rho}\widetilde{C''^2})+\vec\nabla\cdot(\overline{\rho}\widetilde{\vec u}\widetilde{C''^2})&=&
\vec\nabla\cdot\Bigl[\bigl( D+D_{\widetilde{C''^2}}^t\bigr)\overline{\rho}\vec\nabla\widetilde{C''^2}\Bigr]-
\nonumber\\&-&\overline{\rho}\widetilde{\chi}
+2\overline{\rho}D_{\widetilde C}^t(\vec\nabla\widetilde{C})^2+2\overline{\rho}\widetilde{C''\dot\omega''_C},
\end{eqnarray}
where $D$ is the diffusion coefficient for all of the species, given as $D=\nu/Pr$ evaluated assuming a unity Lewis number; $\nu$ is the kinematic viscosity and $Pr$ the Prandtl number; $D_{\widetilde Z}^t=D_{\widetilde{Z^{''2}}}^t=D_{\widetilde C}^t=D_{\widetilde{C''^2}}^t=\nu/Sc_{t}$ are the turbulent mass diffusion coefficients and $Sc_{t}$ the Shmidt turbulent number; $\dot{\omega}_{C}$ is the source for the progress variable. The gradient transport assumption for turbulent fluxes is used and the mean scalar dissipation rate, $\widetilde \chi$, appears as a sink term in equations \eqref{zvar} and~\eqref{cvar}.\\
At every iteration, the values of the flamelet variables of the model are updated and the Favre-averaged thermo-chemical quantities are defined, using equation \eqref{media}. Such solutions provide the mean-mass-fractions which are used to evaluate the flow variables by means of the finite-volume numerical method developed by Cutrone \textit{et al.}~\cite{luigi}. 

\section{NUMERICAL RESULTS}

\begin{figure}
\begin{center}
\subfigure[$Flame\ D$]{\includegraphics[scale=0.2]{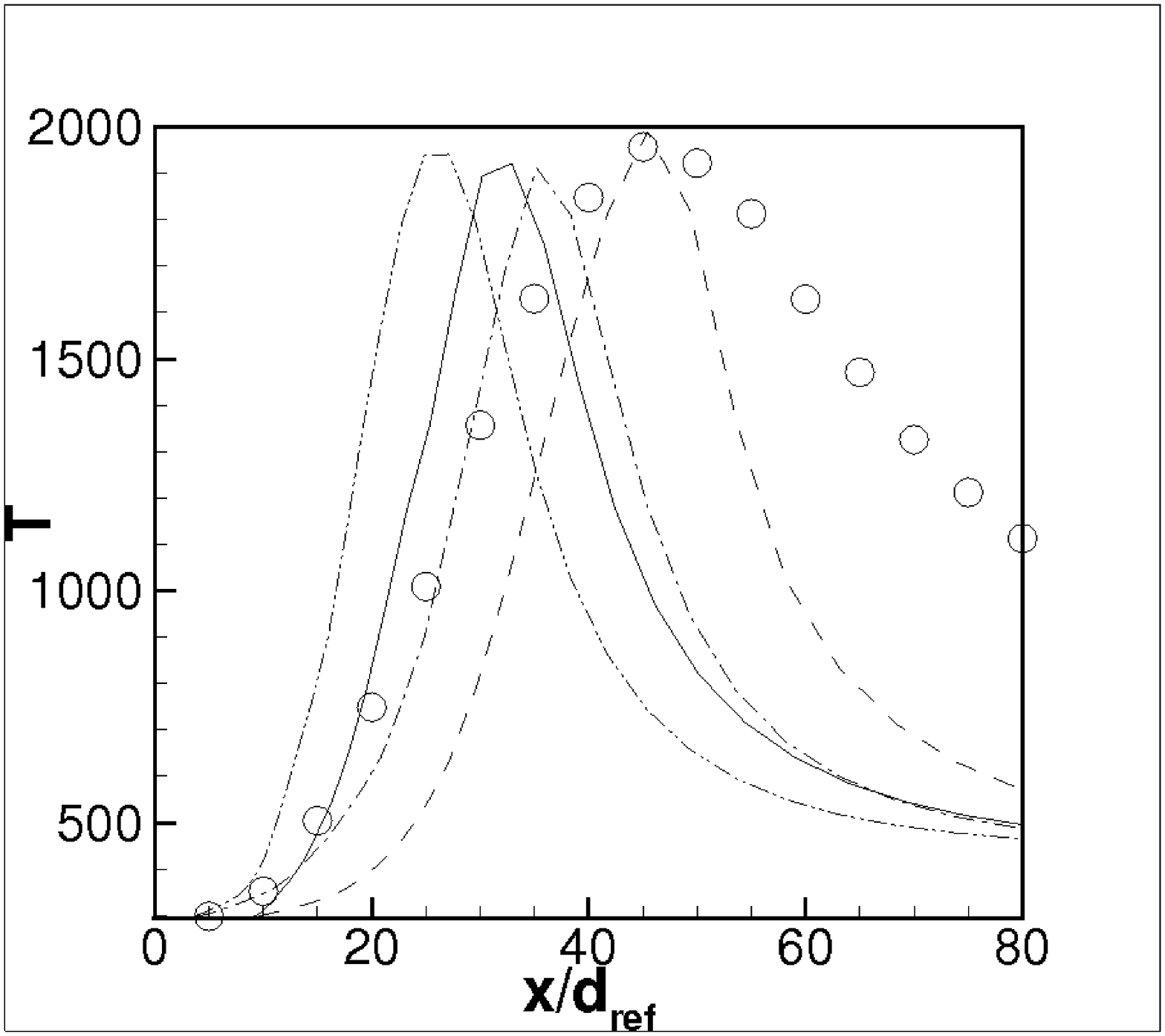}}
\subfigure[$Flame\ E$]{\includegraphics[scale=0.2]{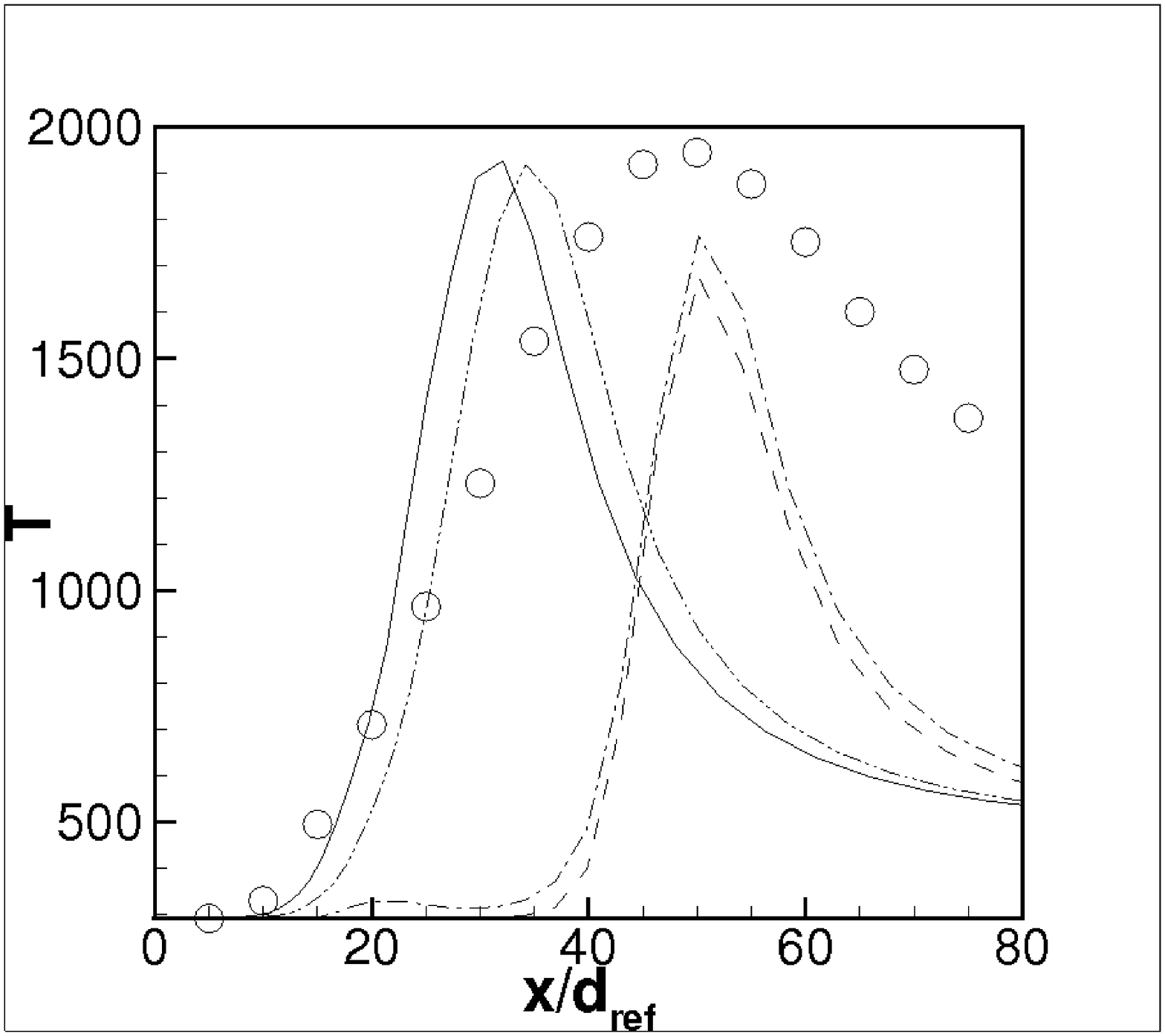}}
\subfigure[$Flame\ F$]{\includegraphics[scale=0.2]{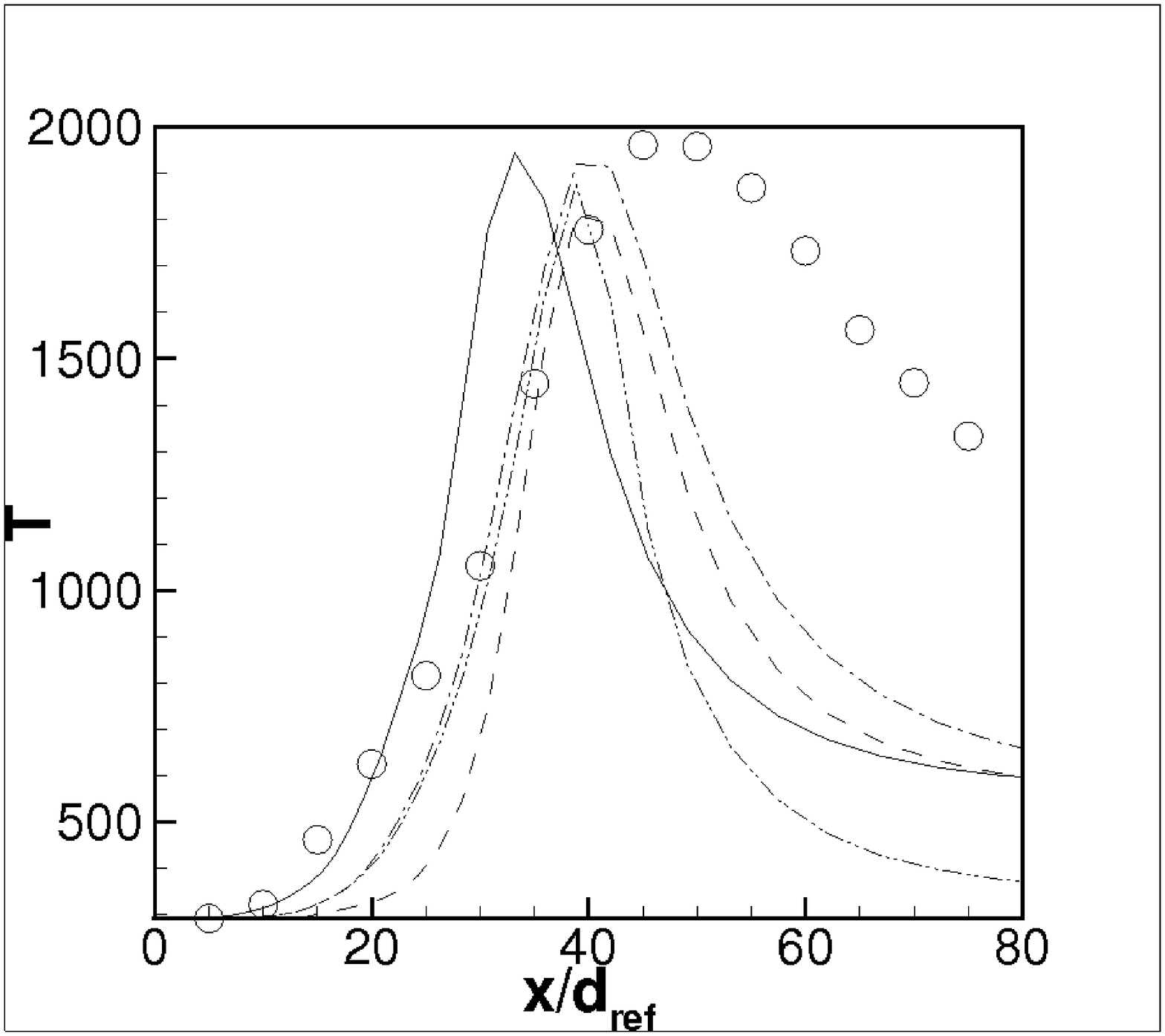}}
\caption{Temperature distribution along the axis of the burner. The solid line is model~D, the long-dashed line is model~C, the dashed-dotted one is model~B and the dashed line is model~A. Symbols are the experimental data.}
\label{temp_center}
\end{center}
\end{figure}

\begin{figure}
\begin{center}
\subfigure[$Flame\ D$]{\includegraphics[scale=0.2]{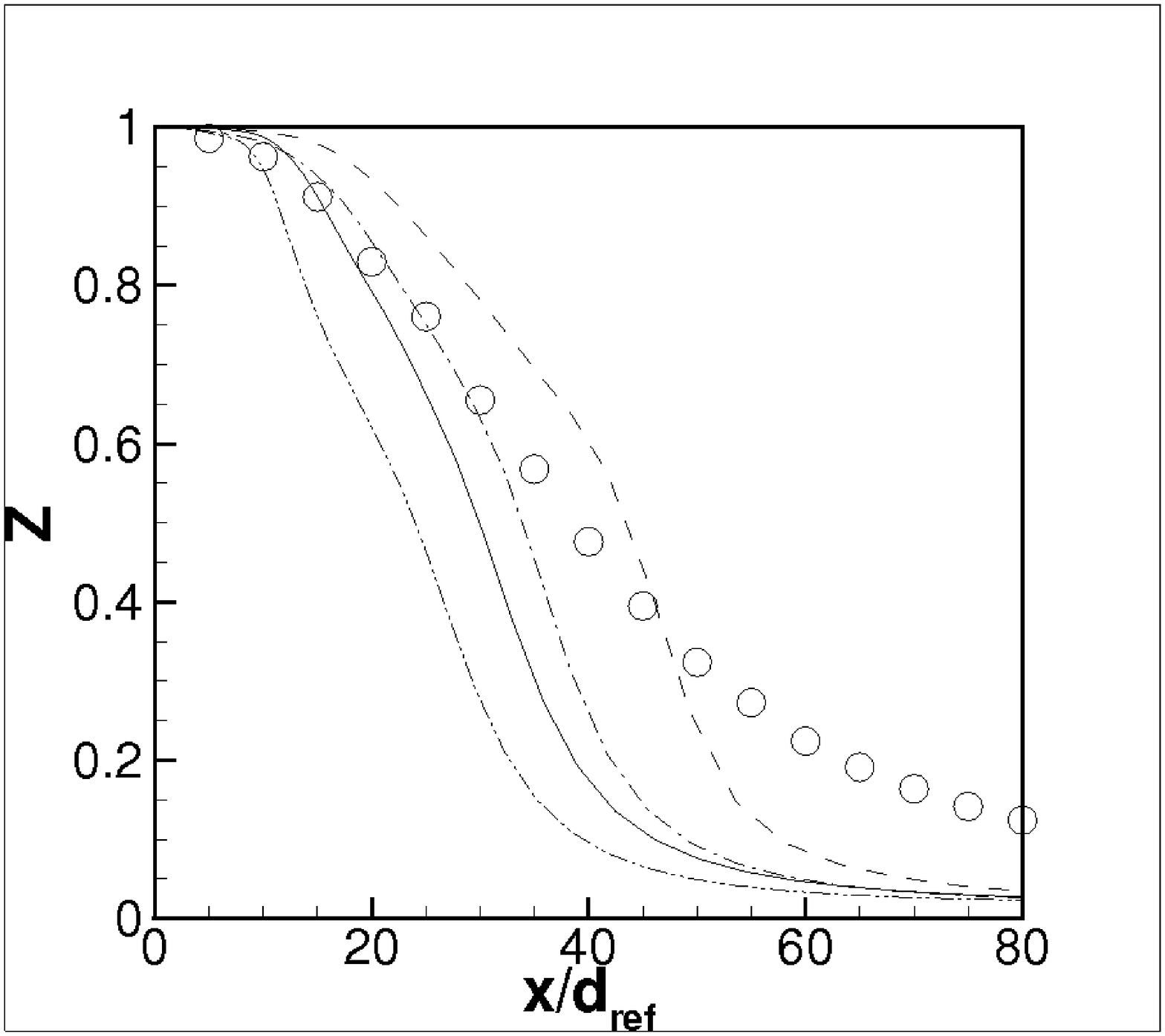}}
\subfigure[$Flame\ E$]{\includegraphics[scale=0.2]{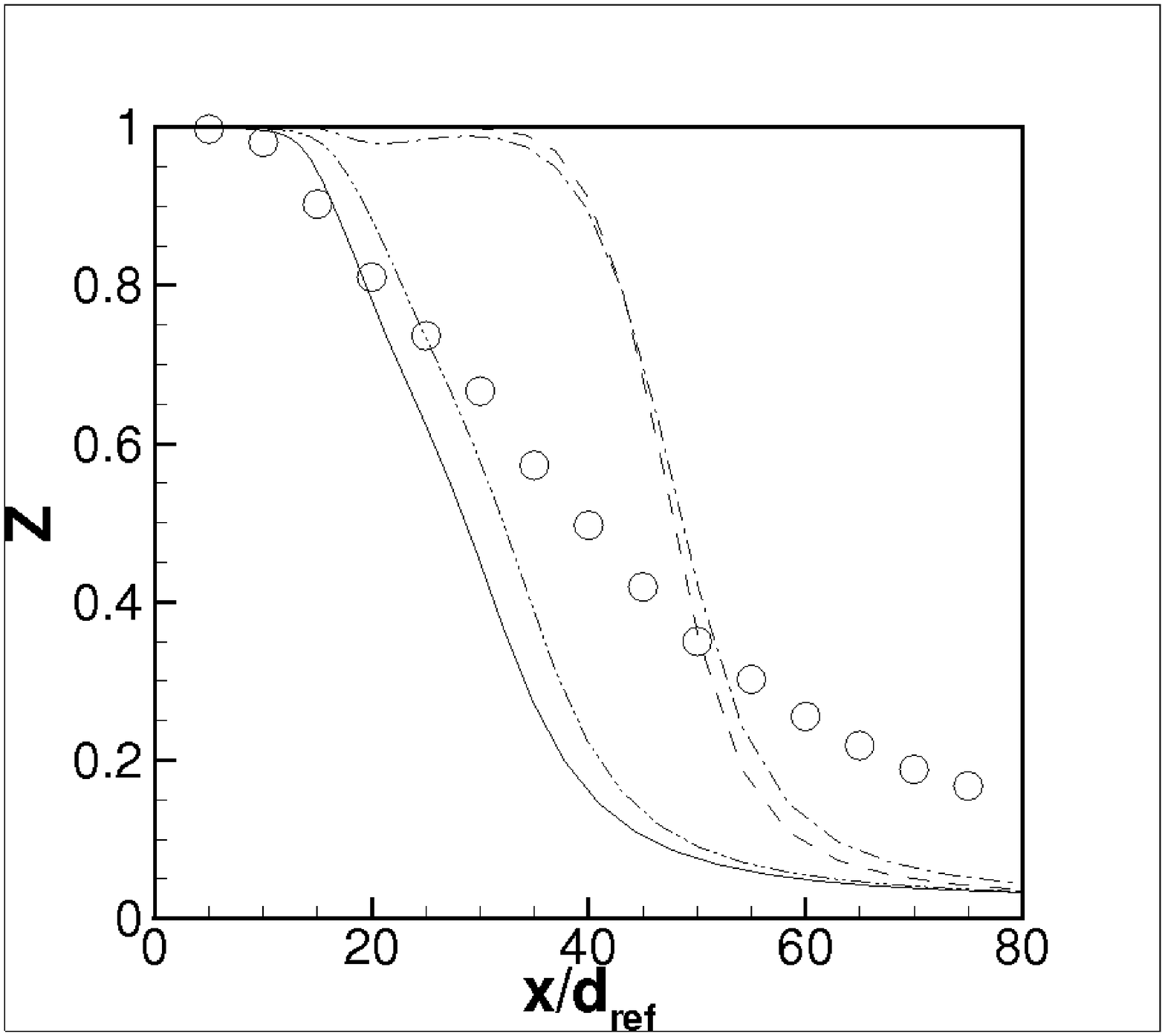}}
\subfigure[$Flame\ F$]{\includegraphics[scale=0.2]{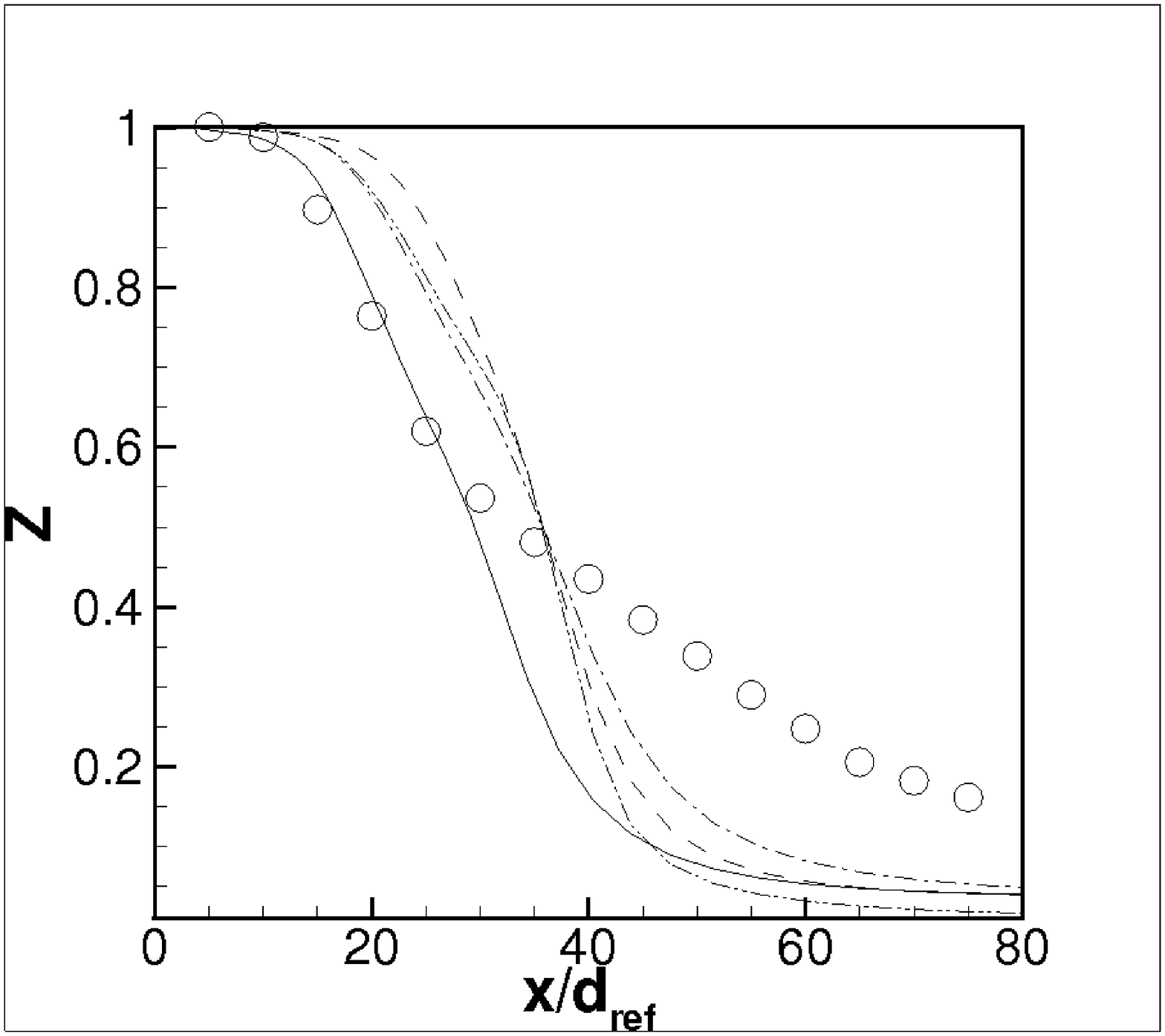}}
\caption{Mixture fraction distribution along the axis of the burner. The solid line is model~D, the long-dashed line is model~C, the dashed-dotted one is model~B and the dashed line is model~A. Symbols are the experimental data.}
\label{zave_center}
\end{center}
\end{figure}
This section provides the comparison among the results obtained using the four combustion models so as to assess the influence of the PDF choice in the prediction of turbulent non-premixed flames. To this purpose, the well knonw subsonic Sandia flames are computed whose experimental data are available in the literature~\cite{sandia}. The steady flamelet evaluations have been solved using the FlameMaster code~\cite{flamemaster}.\\
The Sandia Flames are three different piloted partially premixed methane-air diffusion flames burning at the same pressure, equal to 
$100.6$ kPa, and at three different Reynolds numbers, $R_e$, based on the nozzle diameter, the jet bulk velocity, and the kinematic 
viscosity of the fuel. The diameter of the nozzle of the central jet is $d_{ref}=7.2$ mm and the internal and external diameters of the annular pilot nozzle are equal to  $7.7$ mm and $18.2$ mm, respectively. 
The fluid jet is a mixture of $75\%$ air and $25\%$ methane by volume~\cite{sandia}.  
The pilot is a mixture of air with the main methane combustion products, namely C$_2$H$_2$, H$_2$, CO$_2$ and N$_2$, with the same 
enthalpy at the equivalence ratio $\Phi =0.77$ corresponding to the equilibrium composition $\widetilde Z=0.27,\ \widetilde{Z''^2}=0.0075,\ \widetilde C=1,\ \widetilde{C''^2}=0$. 
The oxidizer air (Y$_{O_2}$=0.233, Y$_{N_2}$=0.767) is supplied as a co-flow at $291$ K.
Flame D ($R_e=22400$) presents very low degree of local extinction, whereas Flame E ($R_e =33600$) and Flame F ($R_e =44800$) have significant and increasing probability of local extinction near the pilot. 
The computational domain is axisymmetric and includes a part of the burner; it has a length of $150\ d_{ref}$ and $27\ d_{ref}$ along the axial and radial directions, respectively, and has been discretized using about $45000$ cells. Computations have been carried out using the combustion scheme described by the GRI-MECH~3.0~\cite{grimech30}: $325$ sub-reactions upon $53$ species. The flamelet library is computed 
over a grid with $125$ uniformly distributed points in the $\widetilde Z$ and $\widetilde C$ directions and $25$ uniformly distributed points in the $\widetilde{Z''^2}$ and $\widetilde{C''^2}$ directions.\\
\begin{figure}[h!]
\begin{center}
\subfigure[]{\includegraphics[scale=0.2]{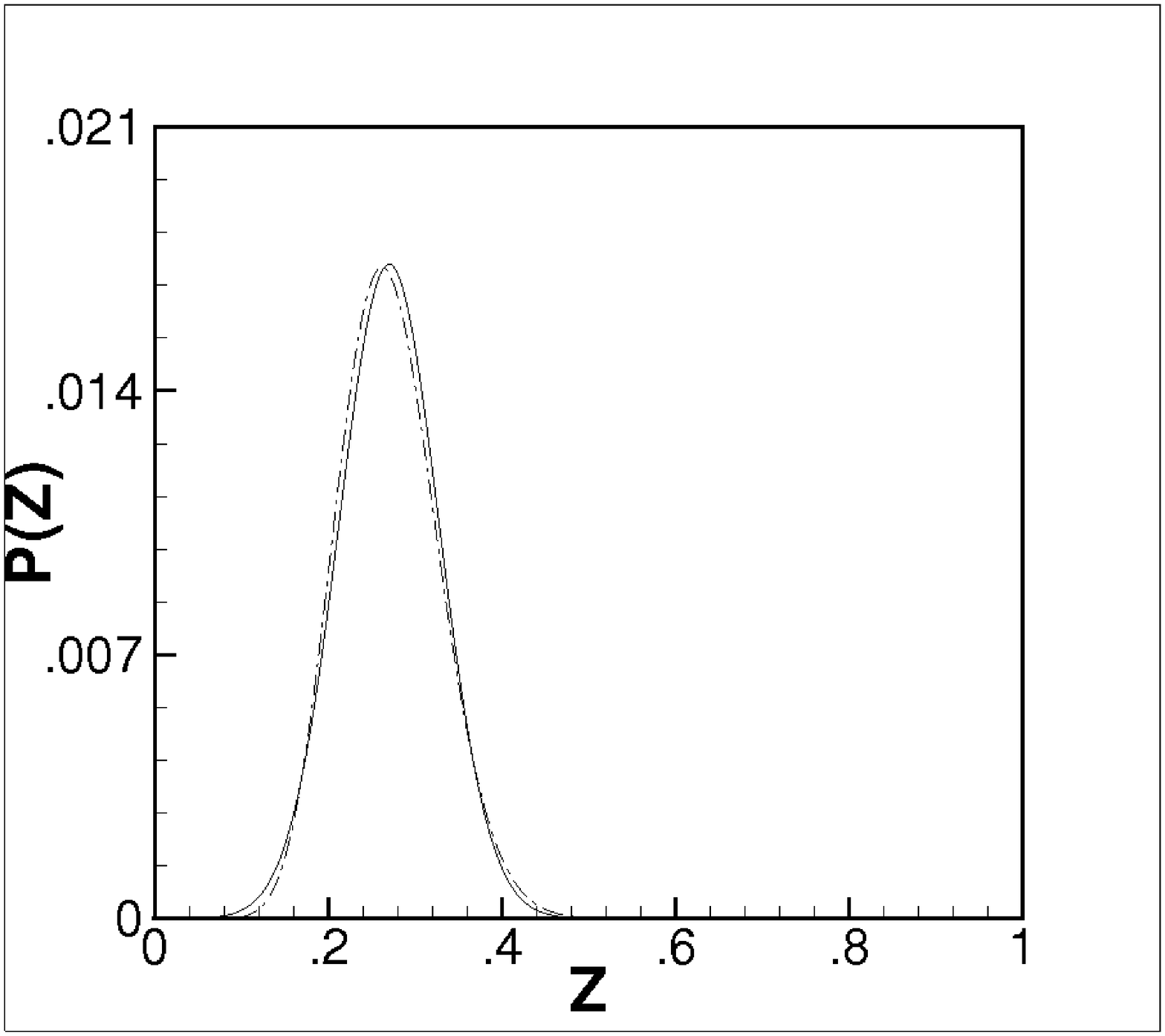}}
\subfigure[]{\includegraphics[scale=0.2]{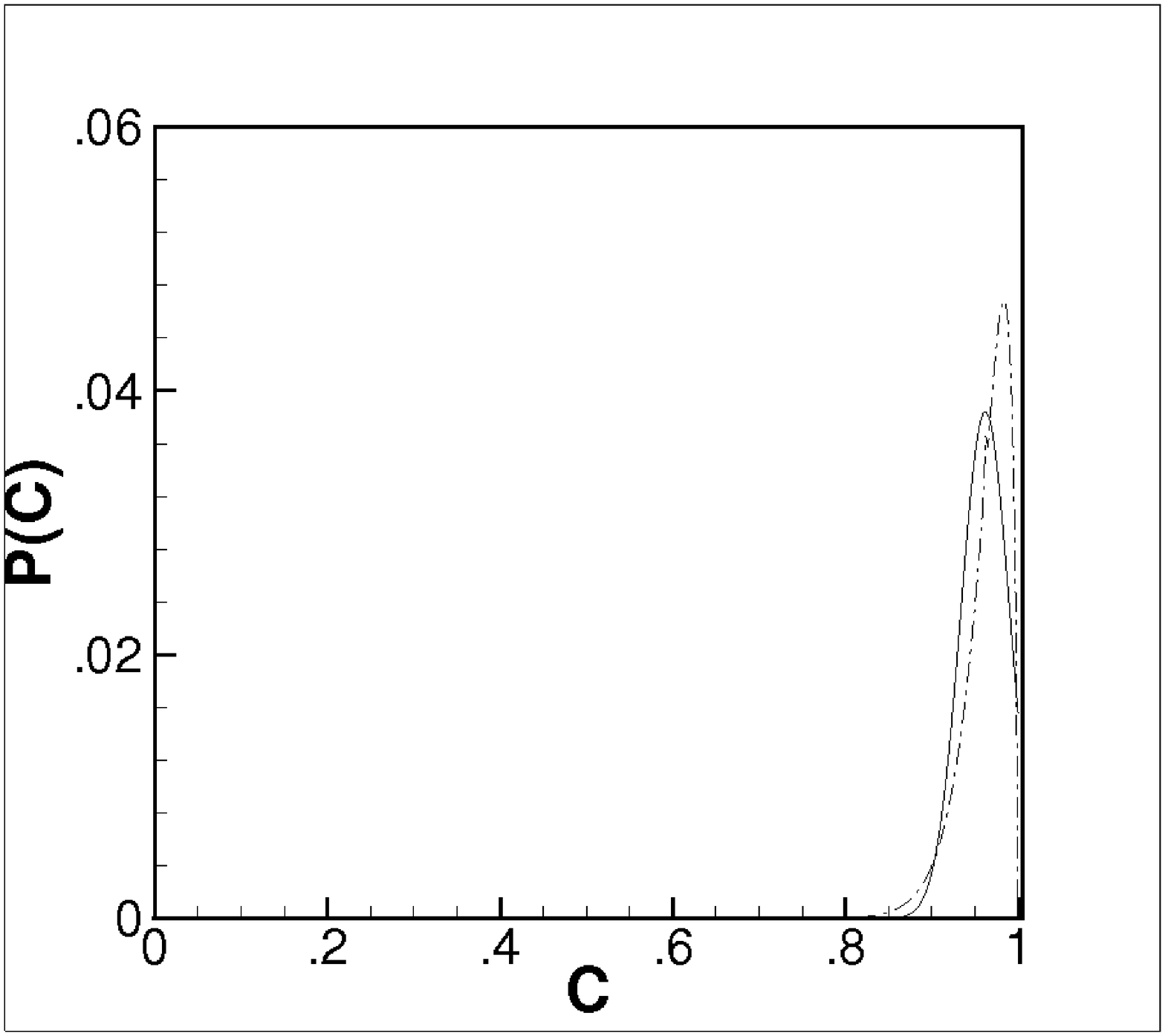}}
\subfigure[]{\includegraphics[scale=0.2]{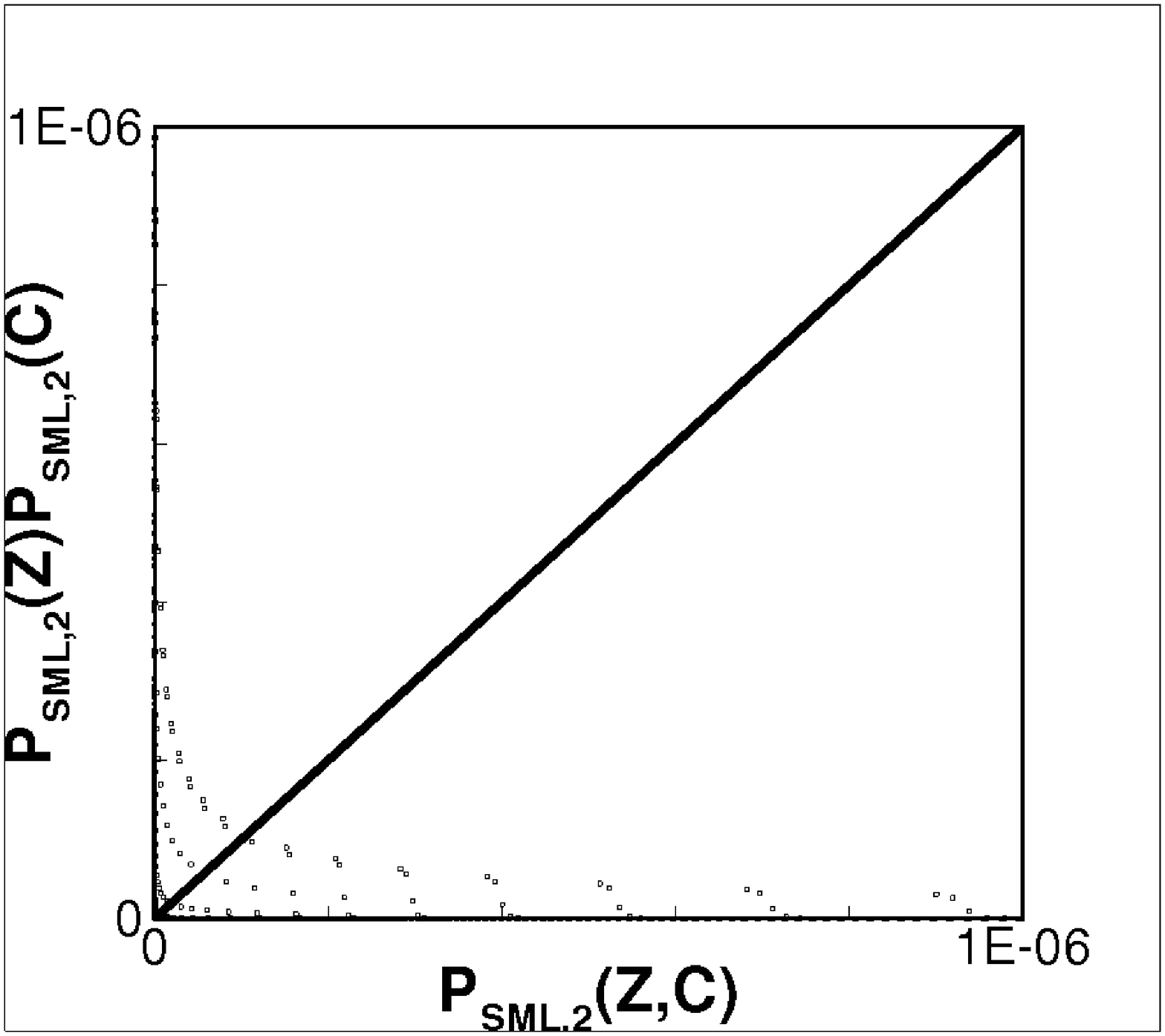}}
\caption{The first two plots, (a) and (b), correspond to the probability distribution of $Z$ and $C$ at the point $(x/d_{ref},y/d_{ref})=(0,1)$. The solid line is the $P_{SML,2}$ distribution and the dashed one the $\beta$-distribution. The (c) plot is the scatter-plot of the joint PDF with statistical independence hypothesis versus the joint PDF, here the solid line is the bisector.}
\label{x_0_y_1}
\end{center}
\end{figure}
Figure~\ref{temp_center} provides the temperature distributions along the axis of the burner. It appears that in the near-burner region model~D is in better agreement with the experimental data than the other three models. Moving away from the burner ($x>20\ d_{ref}$) the agreement deteriorates; this is probably due to the accuracy limits of the RANS approach in the prediction of the mixing process that greatly affects combustion. Moreover, figure~\ref{zave_center} shows the mixture fraction distributions along the axis line.  
From this two set of figures one can see that there is a remarkable improvement, provided by model~D, in the evaluation of the flame core, that is particularly evident in the case of Flame~F. It is interesting to analyse the simulation results at the light of the influence and adequateness of the two widely used simplifying hypotheses: the statistical independence of $Z$ and $C$ and the $\beta$-distribution assumption for $P(Z)$. 
Therefore, some peculiar points have been selected in the computed flow field of Sandia Flame~E simulations and the corresponding values of mean and variance for both $Z$ and $C$ have been used to mark out the distributions. For the first point, with normalized coordinates $(x/d_{ref},y/d_{ref})=(0,1)$  (taken on the burner), the following values are registered: $\widetilde Z=0.2700$, $\widetilde{Z''^2}=0.0034$, $\widetilde C=0.9618$ and $\widetilde{C''^2}=0.0009$. The resulting PDFs are shown in figures~\ref{x_0_y_1}~(a) and~\ref{x_0_y_1}~(b). It appears that the $\beta$-distribution assumption for the mixture fraction and the most likely distribution of $Z$ are in very good agreement. On the other hand, for the distribution of $C$, one can see that the two PDFs are quite different, providing different maximum locations and thus different results in the evaluation of the thermodynamic means. It is interesting to note that since $\beta(Z)$ and $P_{SML,2}(Z)$ are almost coincident, model~C and model~D differ only for the statistical independence hypothesis. This issue is further analysed in figure~\ref{x_0_y_1}~(c), showing a scatter-plot of $P_{SML,2}(Z,C)$ versus the same joint PDF evaluated with the statistical independence hypothesis, namely, $P_{SML,2}(Z)P_{SML,2}(C)$. The reference bisector is also reported as a measure of unitary correlation. It appears that for Flame~E, in the region close to the burner the independence hypothesis is not appropriate.\\
\begin{figure}[h!]
\begin{center}
\subfigure[]{\includegraphics[scale=0.2]{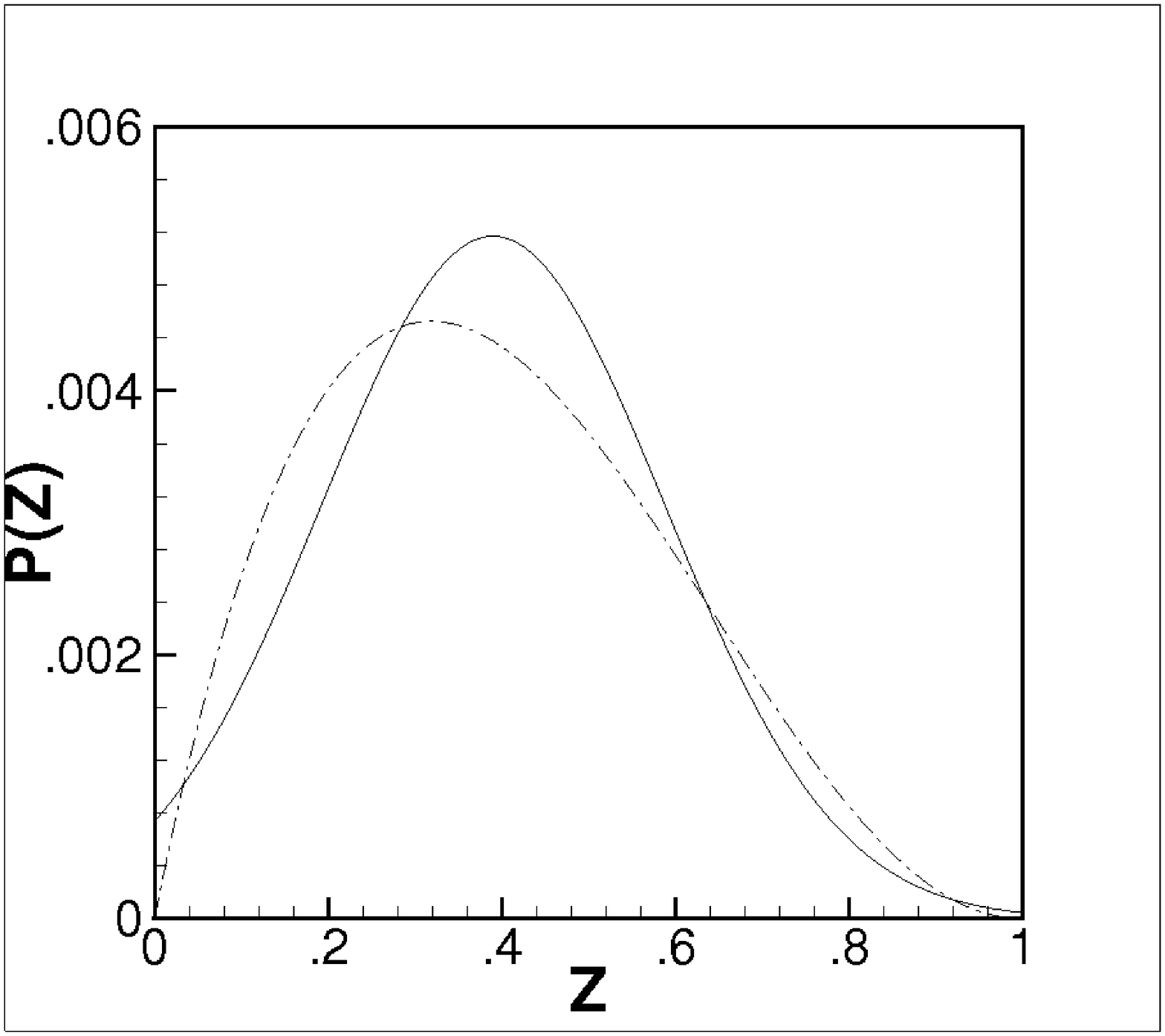}}
\subfigure[]{\includegraphics[scale=0.2]{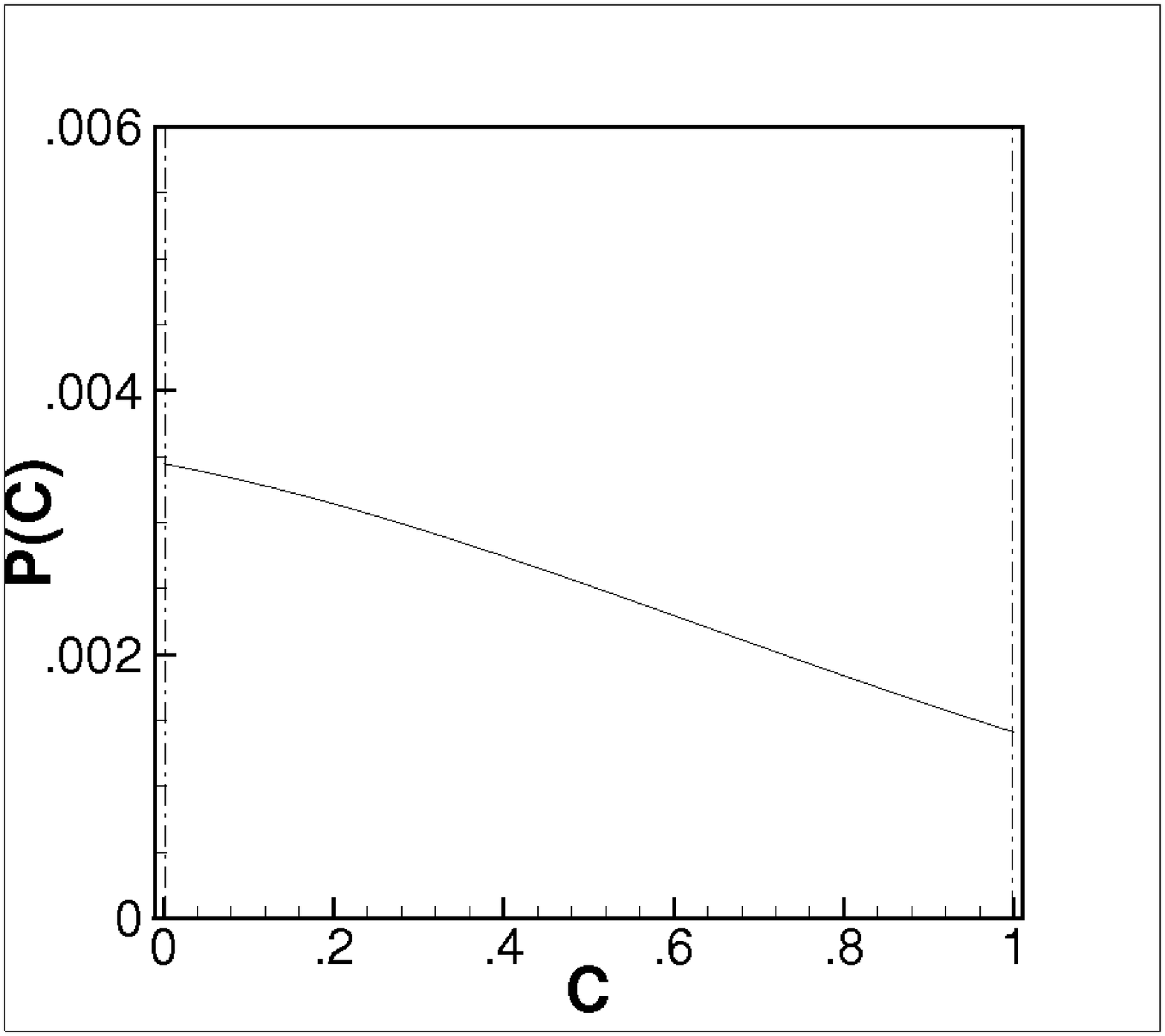}}
\subfigure[]{\includegraphics[scale=0.2]{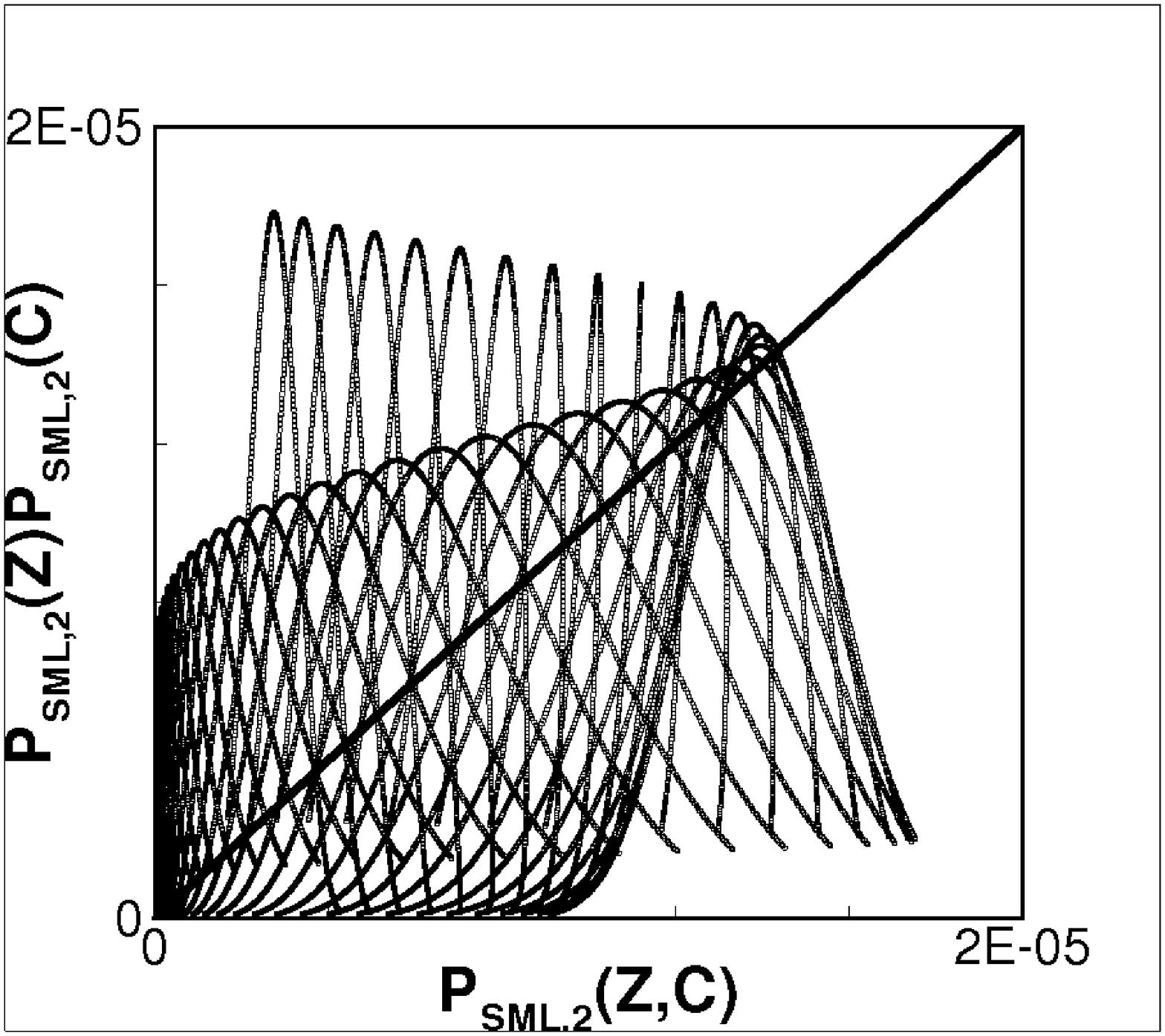}}
\caption{The first two plots, (a) and (b), correspond to the probability distribution of $Z$ and $C$ at the point $(x/d_{ref},y/d_{ref})=(20.85,1)$. The solid line is the $P_{SML,2}$ distribution and the dashed one the $\beta$-distribution. The (c) plot is the scatter-plot of the joint PDF with statistical independence hypothesis versus the joint PDF, here the solid line is the bisector.}
\label{x_20_85_y_1}
\end{center}
\end{figure}
Consider, now, a second point $(x/d_{ref},y/d_{ref})=(20.85,1)$, with the same ordinate, $y/d_{ref}=1$, far from the burner. Here the values of mean and variance are: $\widetilde Z=0.3914$, $\widetilde{Z''^2}=0.0397$, $\widetilde C=0.2074$ and $\widetilde{C''^2}=0.6217$. At this point the hypothesis that the $Z$ is distributed according to a $\beta$-function fails, even if not dramatically, as shown in figure~\ref{x_20_85_y_1}~(a) and~\ref{x_20_85_y_1}~(b). More importantly, in the same figure, one can find that the $P_{SML,2}(C)$ is much more smooth than $\beta(C)$, that has two marked peaks, at $C=0$ and $C=1$. In this case, the statistical independence hypothesis is slightly more appropriate than in the previous case; in fact one can see in figure~\ref{x_20_85_y_1}~(c) that a non negligible part of the points are located near the bisector. Anyway, it still appears an incorrect hypothesis that should be abandoned in order to have an improvement in the combustion simulation. 
 
\section{CONCLUSIONS}
This paper provides an extension of standard FPV model for the simulation of turbulent non-premixed combustion. 
The paper analyses the constitutive hypotheses for the choices adopted in the literature for the presumed PDFs, discussing their adequateness and feasibility.
Then a combustion model is developed with a closure method that, using the SMLD technique, allows one to define the most probable joint PDF of mixture fraction and progress variable. 
The features of the combustion models obtained by the different PDF choices is verified by numerical results obtained for the case of the Sandia Flames. The numerical data are also employed to study the validity of the statistical independence hypothesis. The analysis performed shows that the commonly used hypotheses in the definition of the joint PDF can be discarded in order to have a better estimation of such a PDF; this, in turn, provides a better agreement with experimental data. 
The implementation of the developed model is not expensive since the closure technique is based on an analytical form of the Lagrange's multipliers.
\bibliographystyle{elsarticle-num.bst}
\bibliography{coclite_AIMETA_2013_last.bib}

\end{document}